\documentclass[%
 reprint,
superscriptaddress,
 nofootinbib,
 bibnotes,
 amsmath,amssymb,
 aps,
 longbibliography,  
 prd,
 floatfix,
 aas_macros
]{revtex4-1}

\usepackage{graphicx}
\usepackage{amssymb}
\usepackage{float}
\usepackage{color}
\usepackage{comment}
\usepackage{aas_macros}
\usepackage{soul}
\usepackage[hyperfootnotes=false]{hyperref}
\hypersetup{
     colorlinks = true,
     linkcolor = blue,
     anchorcolor = blue,
     citecolor = blue,
     filecolor = blue,
     urlcolor = blue
     }

\definecolor{todo}{RGB}{228, 86, 73}        
\definecolor{ochre}{RGB}{193, 132, 1}     
\definecolor{review}{RGB}{1, 132, 188}      
\definecolor{done}{RGB}{80, 161, 79}        
\definecolor{question}{RGB}{166, 38, 164}   
\definecolor{comment}{RGB}{79, 82, 94}      
\definecolor{pink}{RGB}{255, 20, 147}      
\definecolor{darkgreen}{RGB}{5, 79, 12}

\newcommand{\skyPosition}{\vec{x}_{\rm s}}
\newcommand{\GWSkymap}{\mathcal{F}_{\rm GW}(\skyPosition)}
\newcommand{\reducedSkymap}{\mathcal{F}^{90}_{\rm GW}(\skyPosition)}
\newcommand{\nupsf}{\mathcal{F}_{\nu}(\skyPosition)}
\newcommand{\GWSignal}{\mathcal{S}_{\rm GW}(\skyPosition)}
\newcommand{\NeutrinoSignal}{\mathcal{S}_{\nu}(\skyPosition)}
\newcommand{\GWBackground}{\mathcal{B}_{\rm GW}}
\newcommand{\NeutrinoBackground}{\mathcal{B}_{\nu}}
\newcommand{\GW}{gravitational wave }
\newcommand{\pipeline}{LLAMA}
\newcommand{\gwastro}{\href{https://gw-astronomy.org/gwhen}{\texttt{gw-astronomy.org}} }
\newcommand{\ninetyCR}{90\% confidence region }
\newcommand{\ninetyCRs}{90\% confidence regions }
\newcommand{\hen}{high energy neutrino }

\newcommand{\gw}{gravitational wave }

\newcommand{\gcn}{GCN}
\newcommand{\gwhen}{the O2 multi-messenger pipeline }
\newcommand{\Gwhen}{The O2 multi-messenger pipeline }
\newcommand{\DAG}{directed acyclic graph }
\newcommand{\etal}{\mbox{et al.}}

\newcommand{\xrays}{\mbox{X-rays}}

\newcommand{\grays}{\mbox{$\gamma$-rays}}
\newcommand{\emready}{\texttt{EM\_READY }}
\newcommand{\columbiaPhysDept}{Department of Physics, Columbia University, New York, NY 10027}
\newcommand{\uFloridaPhysDept}{Department of Physics, University of Florida, Gainesville, FL 32611}
\newcommand{\columbiaAstrophysDept}{Columbia Astrophysics Laboratory, Columbia University, New York, NY 10027}
\newcommand{\desy}{DESY, D-15738 Zeuthen, Germany}
\newcommand{\umd}{Department of Physics, University of Maryland, College Park, MD 20742, USA}
\newcommand{\stockholmUniversity}{Oskar Klein Centre and Department of Physics, Stockholm University, SE-10691 Stockholm, Sweden}

\newcommand*{\TODO}[1]{\textcolor{todo}{\textbf{\textul{TODO}} #1}}

\def\deg{\hbox{$^\circ$}}

\newcommand*{\stefan}[1]{\textcolor{ochre}{\textul{\textbf{\{Stefan\}}} #1}}

\newcommand*{\azadeh}[1]{\textcolor{pink}{\textul{\textbf{\{Azadeh\}}} #1}}

\usepackage{graphicx}
\usepackage{dcolumn}
\usepackage{bm}

\begin{document}

\title{
    Low-Latency Algorithm for Multi-messenger Astrophysics (LLAMA) \\ with Gravitational-Wave and High-Energy Neutrino Candidates
}

\author{Stefan~Countryman}
	\email{Email: stefan.countryman@ligo.org}
	\homepage{http://markalab.org}
	\address{\columbiaPhysDept} 
\author{Azadeh~Keivani}
    \address{\columbiaPhysDept}

\author{Imre~Bartos}
    \address{\columbiaPhysDept}
    \address{\uFloridaPhysDept}
    
\author{Zsuzsa~Marka}
    \address{\columbiaAstrophysDept}
    
\author{Thomas~Kintscher}
	\address{\desy}
    
\author{K.~Rainer~Corley}
    \address{\columbiaPhysDept}
    
\author{Erik~Blaufuss}
	\address{\umd}

\author{Chad~Finley}
    \address{\stockholmUniversity}

\author{Szabolcs~Marka}
    \address{\columbiaPhysDept}

\date{\today}
             

\begin{abstract}
%
We describe in detail the online data analysis pipeline that was used in the multi-messenger search for common sources of gravitational waves (GWs) and high-energy neutrinos (HENs) during the second observing period (O2) of Advanced LIGO and Advanced Virgo. 
Beyond providing added scientific insight into source events, low-latency coincident HENs can offer better localization than GWs alone, allowing for faster electromagnetic follow-up. Transitioning GW+HEN analyses to low-latency, automated pipelines is therefore mission-critical for future multi-messenger efforts. The O2 Low-Latency Algorithm for Multi-messenger Astrophysics (\pipeline) also served as a proof-of-concept for future online GW+HEN searches and led to a codebase that can handle other messengers as well. During O2, the pipeline was used to take LIGO/Virgo GW candidates as triggers and search in realtime for temporally coincident HEN candidates provided by the IceCube Collaboration that fell within the \ninetyCR of the reconstructed GW skymaps. The algorithm used NASA's Gamma-ray Coordinates Network to report coincident alerts to LIGO/Virgo's electromagnetic follow-up partners. 
\end{abstract}


\maketitle



\section{\label{sec:intro}Introduction}

Recent multi-messenger discoveries of GWs in coincidence
with electromagnetic (EM) observations have opened up new windows to the Universe. The multi-messenger science reach of the GW detectors had been enabled by decades of effort preceding the discovery~\cite{LIGOG060660,2006ivoa.spec.1101S,2007AAS...211.7702M,2008APS..APR.E8006A,2008CQGra..25k4039A,2008CQGra..25k4051A,2008HEAD...10.1501M,2009IJMPD..18.1655V,2010AAS...21540604M,2010APS..APRK13004M,2010JPhCS.243a2001M,2011APh....35....1B,2011CQGra..28k4013M,2011GReGr..43..437C,2011ivoa.spec.0711S,2011PhRvL.107y1101B,2012JPhCS.363a2022B,
2012PhRvD..85j3004B,2012PhRvD..86h3007B,2013APh....45...56S,2013CQGra..30l3001B,2013JCAP...06..008A,2013PhRvL.110x1101B,2013RvMP...85.1401A,2014PhRvD..90j1301B,2014PhRvD..90j2002A,2015PhRvL.115w1101B,2016PhRvD..93l2010A,2017ApJ...848L..12A,2017ApJ...850L..35A,2017PhRvD..96b2005A,2017PhRvD..96b3003B,2018cosp...42E2180M}.
The detection of a short gamma-ray burst (sGRB) 1.7 seconds after the GW
detection from a binary neutron star (BNS) merger on August 17, 2017 (GW170817)
was the first multi-messenger/multi-wavelength observation of
GWs~\cite{gw170817,2017ApJ...848L..12A}. The LIGO/Virgo detectors recorded the
GW170817 signal, which was followed by the
detection of GRB\,170817A by \textit{Fermi}-GBM~\cite{GBM-GCN,fermi-gbm} and
\textit{INTEGRAL}~\cite{INTEGRAL-GCN, integral-gw170817},
that were both spatially and temporally coincident with GW170817. This event
was subsequently followed up by several different observatories in a broad
range of wavelengths and cosmic messengers~\cite{2017ApJ...848L..12A,2017ApJ...850L..35A}. 

Another recent multi-messenger discovery is related to the detection of a
high-energy neutrino (HEN; IceCube-170922A) by the IceCube Neutrino Observatory,
which is the first $3\sigma$ correlation with EM emissions from a flaring
blazar, TXS 0506+056~\citep{ic1709022mm,keivani2018}. 
The flaring blazar was in an active phase in high-energy and very high-energy
\grays\ and showed variabilities in \xrays\ and radio bands.
This confirms that HENs are produced by cosmic accelerators such as blazars. 

These two recent multi-messenger discoveries enable us to better understand and
explore the origin of cosmic particles, the astrophysical mechanisms that
produce them, and the physical implications of their sources.

The detection of
GWs~\cite{gw150914,gw151226,gw170104,gw170608,gw170814,gw170817} and
HENs~\cite{ic-astronu14,ic-science13,ic_icrc17} have been separately reported
and confirmed over a few years of detector operations, though no astrophysical
source has yet been observed simultaneously with both
messengers~\cite{2011PhRvL.107y1101B,2016PhRvD..93l2010A,2017PhRvD..96b2005A,2017ApJ...850L..35A,2018arXiv181010693A}. 
Observing even a single joint source of GWs and HENs in both messengers could
transform our understanding of the underlying mechanisms that create
them~\cite{2013RvMP...85.1401A,2013CQGra..30l3001B}.

A previous paper by Baret \etal\ \citep{2012PhRvD..85j3004B} presented a joint GW+HEN
analysis algorithm that has been used for previous searches and which was
adapted for the Low-Latency Algorithm for Multi-Messenger Astrophysics (\pipeline).
It was used in a joint search using Initial LIGO/Initial Virgo
and the partially completed IceCube detector data, which placed upper limits on the
source rate~\cite{2014PhRvD..90j2002A}. The algorithm described below builds on this
work. There had also been earlier efforts on designing joint searches for GWs
and HENs, such as~\citep{2008CQGra..25k4039A} and \citep{2009NIMPA.602..268P}.

The rapid EM follow-up of GW events from compact binary mergers has been made
significantly easier by LIGO/Virgo's low latency alert
distribution to partner
observatories~\citep{messick2017,gbm-gw170817,integral-gw170817}.
However, the typically large
source localization
uncertainty from GW data creates challenges for EM follow-ups due to EM telescopes'
relatively small fields of view. Another problem with limited localization is
the large number of foreground transients, mostly supernovae, that are
cumbersome to differentiate from GW
counterparts~\cite{bartos2015,Cowperthwaite2017}.
By contrast, HENs from IceCube provide localizations of a median of $\sim 0.5
\deg$ for an $E^{-2}$ signal neutrino spectrum~\citep{icgfu2016}. This, in
addition to high duty cycle of neutrino observatories, makes HENs also a
well-suited messenger to enhance GW studies.
The rapid identification of GW+HEN
coincidences will provide significantly more precise localization than GWs
alone, enabling faster and more efficient EM follow-up observations with a
high scientific payoff.

Several sources capable of generating HENs and GWs have been proposed,
including core-collapse supernovae
(CCSN)~\cite{2012PhRvD..86h3007B, kohta-sn-2018}, GRBs~(see e.g.~\cite{kohta2006,meszaros13}),
BNS mergers~\cite{2018arXiv180511613K},
neutron star-black hole
(NS-BH) mergers~\cite{shigeo2017}, soft gamma
repeaters~\cite{ioka2005, mereghetti2008}, 
and microquasars~\cite{2009NIMPA.602..268P}.
Core-collapse supernovae have long been considered sources of GW emission
(e.g.~\citep{2012PhRvD..86h3007B,2004PhRvL..93r1101R,2005PhRvL..95f1103A}).
These sources create relativistic outflows
capable of producing HENs that can travel unimpeded for billions of light
years before reaching Earth, providing useful information when they interact
with the detectors.
TeV neutrinos are also expected to be detected in the HEN detectors from a
galactic SN $\sim 0.1-10$ days after detections of GWs and MeV
neutrinos~\cite{kohta-sn-2018}.

Among the most promising multi-messenger sources are the progenitors of GRBs.
sGRBs associated with BNS (or NS-BH) mergers are known to be GW
emitters~\cite{2017ApJ...848L..12A} and are considered to be HEN
sources~\cite{meszaros13}. Binary Black Hole (BBH) mergers emitting
GWs (e.g.~\cite{gw150914,gw151226}) are typically not expected to be strong
emitters of EM and neutrino radiation, although there could be exceptions \cite{2017ApJ...835..165B,2016ApJ...822L...9M,2017MNRAS.464..946S,2016ApJ...821L..18P,2017NatCo...8..831B}.

Besides the known sources, searching for astrophysical GW+HEN signal might
reveal unknown sources or production mechanisms. 

This paper describes \pipeline\ in the configuration that searched for joint GW+HEN
sources from LIGO/Virgo during LIGO's second observing run (O2).
In Sec.~\ref{sec:detectors} we describe the state of the LIGO and Virgo
detectors as well as the IceCube Neutrino Observatory during O2. We then
summarize past offline searches in Sec.~\ref{sec:past}. In
sec.~\ref{sec:pipeline} we explain the details of the GW+HEN online search,
including the data analysis method and software implementation. We
conclude in Sec.~\ref{sec:conclusions} with lessons learned from the GW+HEN
pipeline in O2.


\section{\label{sec:detectors} Detectors and Data}


\begin{figure*}
    \includegraphics[width=\textwidth]{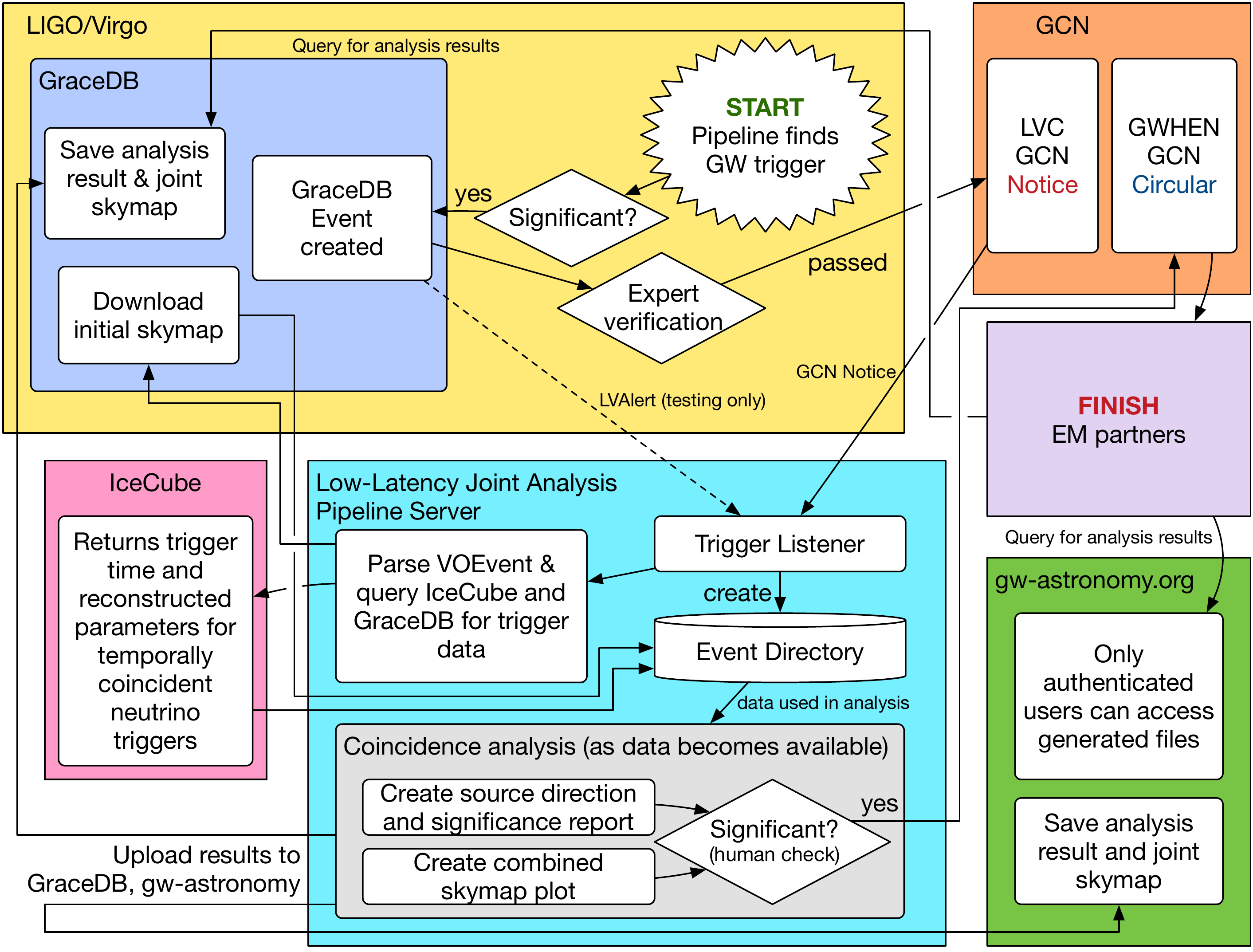}
    \caption{
        Information flow in the pipeline. This diagram shows trigger and
        data sources and destinations used during O2, though the pipeline
        itself can readily accomodate new sources and destinations for both.
        See Sec.~\ref{subsec:flow} for more information.
    }
    \label{fig:flowchart}
\end{figure*}

During O2, \pipeline\ combined data from LIGO, Virgo, and
IceCube, using NASA's Gamma-Ray Coordinates Network (GCN; \cite{gcn}) as a
mediator to receive
triggers and disseminate results. 
Results were stored on LIGO/Virgo's candidate
database (GraceDB\footnote{https://gracedb.ligo.org}) 
and on \gwastro.
Sec.~\ref{subsec:flow} describes this flow of data between partners
(summarized in Fig.~\ref{fig:flowchart}).

\subsection{LIGO and Virgo}
\label{subsec:aligoadvirgo}

LIGO consists of two interferometers, each with 4-km-long
arms. One site is in Livingston, Louisiana and the other is in Hanford,
Washington~\cite{ligo2009}. Virgo is also a similar interferometer located
near Pisa, Italy~\cite{virgo2012}. The arm length of Virgo is 3 kilometers.
Both LIGO and Virgo are sensitive to GWs in a frequency band of
10-10,000 Hz~\cite{aligo-sensitivity}.
The
Initial LIGO upgrades started in 2010 resulted in operational commencement of
LIGO~\cite{aligo2015} in 2015.
LIGO underwent a series of upgrades
between its first and second observing run (O1 and O2, respectively; \cite{lrr2018}).

The second observing run (O2) of the LIGO detectors started in November 30,
2016, with Virgo~\cite{avirgo2015} joining on August 1, 2017. O2 finished on
August 25, 2017. Four GW discoveries were reported from this period:
GW170104~\citep{gw170104}, GW170608~\citep{gw170608},
GW170814~\citep{gw170814}, and GW170817~\citep{gw170817}.
The first three were
emitted from BBH mergers and the last one as mentioned in Sec.~\ref{sec:intro}
resulted from a BNS merger. Triggers from all LIGO/Virgo search pipelines (including
cWB~\cite{2008CQGra..25k4029K,2016PhRvD..93d2004K}, GSTLAL~\cite{messick2017}, PyCBC~\cite{2016CQGra..33u5004U}, oLIB~\cite{2015arXiv151105955L}, and MBTAOnline~\cite{2016CQGra..33q5012A}) were included in the search as long
as they had
been manually approved for EM follow-up by LIGO/Virgo experts
(indicated by GraceDB's \emready flag)
and had had a GCN Notice sent out (acting as a redundant check of 
the event's validity). 
LIGO/Virgo have recently announced a new catalog of GWs including four additional GW candidates detected during O2~\cite{2018arXiv181112907T}.

\subsection{IceCube Neutrino Observatory}
\label{subsec:ic}
IceCube is a gigaton neutrino detector located under the geographical South
Pole in Antarctica~\cite{icecube2017}. IceCube has been continuously
detecting neutrinos with its complete 86-string configuration since
2010~\cite{icecube-7yr-ps}.  


IceCube HENs interact with the ice through
neutral-current and charged-current interactions.
In a charged-current interaction, most of the neutrino energy is transferred to its associated leptons. 
Muons are capable of traveling several kilometers
(unlike electrons, which lose energy rapidly, and taus, which decay quickly),
emitting Cherenkov light along their path that is detected by the IceCube digital
optical modules (DOMs) placed in the polar ice~\cite{icecube2017}.
The direction of the
muon and hence its parent neutrino is reconstructed using the timing
information of photons captured by different DOMs along the muon track.
The reconstruction methods cannot distinguish between neutrinos and anti-neutrinos and are sensitive to their combined flux.
IceCube can identify the direction of these through-going muons with a
precision of $\sim 0.5^\circ$~\cite{icecube-realtime}. The IceCube Collaboration provided us with a stream of realtime muon data~\cite{icecube-realtime} for testing this pipeline.

The primary background source in IceCube comes from the muons that are produced
by cosmic ray interactions in the atmosphere~\cite{icecube2017}. 
The up-going events are more likely to be from neutrino interactions, as the Earth attenuates the cosmic-ray induced atmospheric muons.
The mis-reconstructed muons are the main backgrounds left in the northern
hemisphere. 
IceCube uses multivariate selection techniques (Boosted
Decision Tree, BDT; \cite{BDT})
to remove the poorly reconstructed events to distinguish
the signal (muon products from neutrino interactions in ice) from the
background~\cite{icgfu2016,ic-bdt-2014,icecube-7yr-ps}. The other source of
all-sky background consists of atmospheric neutrinos which follow a softer
energy spectrum compared to the typically expected astrophysical neutrino
spectrum. 

In the southern sky, the background is dominated by the atmospheric cosmic-ray
muons which are well-reconstructed. Therefore more restrictive cuts are applied
to reduce the background in the southern hemisphere. In this study, in addition
to the upgoing neutrinos, we accept the down-going events with a BDT score
greater than 0.1~\cite{ic-bdt-2014}.

\subsection{Data Flow to and from the Pipeline}
\label{subsec:flow}

\pipeline\ relied on LIGO/Virgo and IceCube data for input; GCN (as an
intermediary) for GW triggers; LIGO/Virgo and \gwastro for data result
storage; and GCN for dissemination of results via GCN Circular to EM
followup partners (see Fig.~\ref{fig:flowchart} for a diagram of data).

\pipeline\ used GW triggers deemed significant by LIGO/Virgo as
input.
These triggers were generated by LIGO/Virgo detection pipelines
(Sec.~\ref{subsec:aligoadvirgo}), stored on GraceDB along with reconstructed
skymaps, and human-vetted (Sec.~\ref{subsec:aligoadvirgo}). If deemed significant,
they were sent in the form of a VOEvent~\cite{voevent} to GCN and distributed as
a GCN Notice. The pipeline received and parsed these GCN Notices, used
the metadata they contained to fetch GW and neutrino localizations from
LIGO/Virgo and IceCube, and used those localizations to run the
joint analysis (see Sec.~\ref{subsec:method} for a description of the
analysis method and Sec.~\ref{subsec:implementation} for details on the
software implementation of the internal part of the analysis).

Data products from the joint analysis (in the form of a joint skymap and
neutrino candidate data) were uploaded to GraceDB and \gwastro to facilitate
data archiving and access by EM followup partners. A draft GW+HEN GCN Circular
was automatically generated by the pipeline and distributed for verification and (if necessary) modification. Once it
was deemed ready for distribution, it was distributed as a GW+HEN GCN Circular following the GW trigger's LIGO/Virgo
GCN Circular; this final step was a necessary step to avoid confusion
among EM followup partners unfamiliar with \pipeline.

\section{\label{sec:past}Past Offline Searches}

An offline GW+HEN search was conducted for Initial LIGO and Initial
Virgo using HENs from the partially completed IceCube
\cite{2014PhRvD..90j2002A}. The joint search ran during Initial LIGO's 5th and 6th observing
runs (S5 and S6, respectively) and Initial Virgo's first three observing runs (SR1,
SR2, and SR3); IceCube ran in its 22, 59, and 79-string configurations. A joint
search was also conducted with Initial LIGO S5/Initial Virgo SR1 and ANTARES in its 5-string
configuration \cite{2013JCAP...06..008A}. The complete configuration of ANTARES
with data in 2009-2010 coincided with Initial Virgo SR2-3 and Initial LIGO S6 runs and
significantly improved the GW+HEN search
sensitivity~\cite{antares-ligo-ICRC15}. These searches used sub-threshold
GW event candidates (i.e. triggers whose significance 
was too
low for LIGO/Virgo to claim a detection)
from the entire run and calculated a joint test statistic, taking
advantage of the improved sensitivity of a joint search to GW+HEN events. No
significant coincident events were found in either search, leading to upper
limits on joint GW+HEN event rates.

HEN search results and GW+HEN science foundations have also been published for several GW
detections from LIGO's 1st and 2nd observing runs (O1 and O2, respectively)~\cite{2006ivoa.spec.1101S,2007AAS...211.7702M,2008APS..APR.E8006A,2008CQGra..25k4039A,2008CQGra..25k4051A,2008HEAD...10.1501M,2009IJMPD..18.1655V,2010AAS...21540604M,2010APS..APRK13004M,2010JPhCS.243a2001M,2011APh....35....1B,2011CQGra..28k4013M,2011GReGr..43..437C,2011ivoa.spec.0711S,2011PhRvL.107y1101B,2012JPhCS.363a2022B,
2012PhRvD..85j3004B,2012PhRvD..86h3007B,2013APh....45...56S,2013CQGra..30l3001B,2013JCAP...06..008A,2013PhRvL.110x1101B,2013RvMP...85.1401A,2014PhRvD..90j1301B,2014PhRvD..90j2002A,2015PhRvL.115w1101B,2016PhRvD..93l2010A,2017ApJ...848L..12A,2017ApJ...850L..35A,2017PhRvD..96b2005A,2017PhRvD..96b3003B,2018cosp...42E2180M}
and Virgo's first observing run. HEN candidates for these searches were
provided by IceCube, and ANTARES. Data from the Pierre Auger Observatory was
also included in the case of the first detected BNS merger,
GW170817~\cite{2017ApJ...850L..35A}. Other analyzed events were BBH mergers
GW150914~\cite{2016PhRvD..93l2010A}, GW151226 and LVT151012~\cite{2017PhRvD..96b2005A}.
These searches ran in response to high-significance GW candidates (all but LVT151012 were claimed as detections); 
full searches for O1 that include sub-threshold GW events are reported in~\cite{2018arXiv181010693A}.
Full searches for O2 are ongoing.





\section{\label{sec:pipeline}GW+HEN Online Search}

The low-latency joint GW+HEN event search was enabled in response to new event candidates. The rapid response and low-latency analysis expanded capabilities of GW+HEN searches compared to previous archival searches. 
Beyond its obvious discovery potential, a low-latency
search also offers numerous advantages for EM follow-up, including improved
localization (Sec. \ref{subsubsec:localization}), low-latency sub-threshold source
detection (Sec. \ref{subsubsec:sub-threshold}), and the ability to handle
increased event rates through automation
(Sec. \ref{subsubsec:automation}). Achieving these goals
within O2 constraints required nominal analysis times of less than 30 minutes
(Sec. \ref{subsec:timeline}) to run the data
analysis procedure described in Sec.~\ref{subsec:method}. The implementation of \pipeline, described in Sec.~\ref{subsec:implementation}, achieved
this goal, providing joint GW+HEN localization and testing
key methods for future low-latency sub-threshold searches and the planned
full automation of GW+HEN joint searches.

\begin{figure}
    \begin{center}
        \resizebox{0.48\textwidth}{!}{\includegraphics{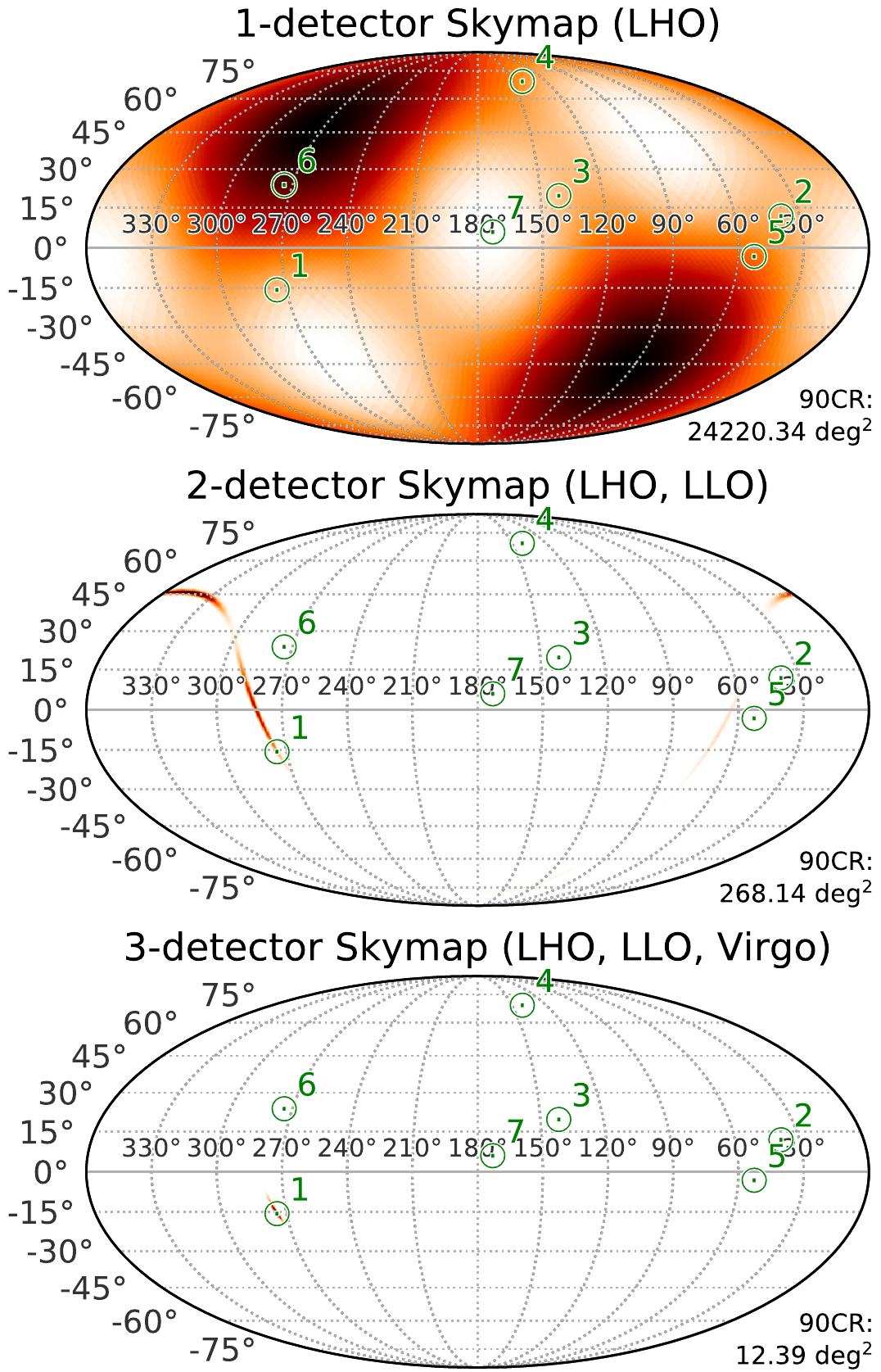}}
    \end{center}
    \caption{
        \textbf{Simulated} joint skymaps showing what a joint
        detection would look like for 1, 2, and 3 detector skymaps from a BNS
        merger with LIGO/Virgo at design sensitivity. The brightest regions
        of the
        skymaps are the likeliest GW source directions; sizes of the 90\%
        credible regions are noted at bottom right.
        Both progenitors have masses
        of $1.4 M_{\odot}$. The neutrinos are located at the green dots
        (surrounding circles added for emphasis). A
        fake coincident neutrino has been inserted into the figures as Neutrino
        1 to show what the skymap would have looked like for a
        coincident detection. Though the neutrino localization is better in all
        three cases, the improvement provided by a joint localization is
        much greater in the 2-detector skymap. 1-detector localization is too
        poor to pick out the correct neutrino from direction alone, though other
        factors (like neutrino time/energy or other non-GW+HEN observations, e.g.
        GRBs) can boost the joint significance
        enough to make a given neutrino trigger worth investigating.
    }
    \label{fig:joint-skymaps}
\end{figure}

\subsection{\pipeline\ Advantages for EM Follow-ups}
\subsubsection{Improved localization with neutrinos}
\label{subsubsec:localization}

The GW search area size is a limiting
and costly
factor in the speed of EM follow-up
efforts for all but the highest energy photons~\cite{2014MNRAS.443..738B, 2018MNRAS.477..639B}. 
Typical EM observatories have viewing areas smaller than 10
deg$^2$, whereas LIGO/Virgo GW skymaps typically have
\ninetyCRs ranging from tens to thousands of
deg$^2$ (depending on the number of detectors included in a trigger and the loudness
of the GW signal). IceCube neutrinos, by contrast, typically have error regions
of 0.5\,deg$^2$. A joint GW+neutrino signal's localization will be determined by better localized messenger, i.e. neutrinos. The typical 0.5\,deg$^2$ localization makes these joint events easier to follow up, particularly when the original GW localization was poor (Fig. \ref{fig:joint-skymaps}).

A low-latency GW+HEN search can identify neutrinos that come from the GW source
and provide their localizations to EM partners, allowing for rapid EM source
identification without having to tile large areas of the sky. This is
particularly important in cases when only one or two GW detectors are
operational, as the typical \ninetyCR will typically be hundreds to thousands of square
degrees in these cases. 1 or 2 detector triggers are a likely scenario given
typical LIGO/Virgo duty cycles of 50-70\%~(\cite{messick2017} and references therein).

Even when all detectors are operational, issues with data transmission and
glitch removal (among other things) can prevent all operating GW detectors from
providing data for a low-latency skymap. This specific scenario occurred during
the detection of GW170817, the first direct detection of a BNS merger; a glitch
in the LIGO Livingston observatory and an issue with Virgo data transmission
caused the earliest available skymap to have 1-detector localization, with
improved localizations (with 2 and 3 detectors)
following later on
as data became available.
Had a neutrino been detected from the merger by IceCube,
\pipeline\ would have provided a faster improved localization than that of the
GW/GRB skymaps alone. Providing rapid joint localization was a primary goal of
the GW+HEN pipeline during O2.

Fig. \ref{fig:joint-skymaps}
shows how the number of contributing GW detectors affects source localization
and how a joint GW+HEN detection can provide much faster source-direction
recovery than a GW skymap alone.

\subsubsection{Low-latency sub-threshold search}
\label{subsubsec:sub-threshold}

\pipeline\ calculates the joint significance of an event from
the significances of the individual triggers. The coincident events 
are expected to
achieve higher significances than the individual events.
Sub-threshold triggers, which, alone have lower significances (in comparison
with the rare high significance events) are by themselves unrecoverable as
astrophysical GW
signals but can sometimes be identified in the company of another
sub-threshold signal in a different messenger channel. This means that
sub-threshold GW and neutrino triggers can be run through a joint analysis with
the intention of finding events whose significances are high enough that
the resulting multi-messenger event candidates exceed the detection threshold.

Past sub-threshold search efforts have been offline, taking place months after
the triggers were identified and thus precluding any chance at prompt EM
follow-up. An online sub-threshold search could identify joint events that
would otherwise not be broadcast to the EM follow-up community because of low
significance of the GW trigger alone. The O2 GW+HEN pipeline is capable of
searching for sub-threshold GW+HEN events in realtime, although this feature
was not utilized during O2 due to 
data use restrictions.
(It was successfully tested internally, however.)

\subsubsection{Automation needed for higher event rate}
\label{subsubsec:automation}

Event rates for both GW and neutrino searches are expected to climb in the
coming years~\cite{lrr2018,2015arXiv151005228T}. 
In addition, sub-threshold searches must include significantly
higher numbers of triggers than standard joint searches. As analysis methods
and observational data APIs converge and stabilize, it
becomes feasible to automate and speed up
these searches and thus avoid analysis backlogs.

\subsection{Target Timeline for an Online Search}
\label{subsec:timeline}

\begin{figure*}
    \includegraphics[width=\textwidth]{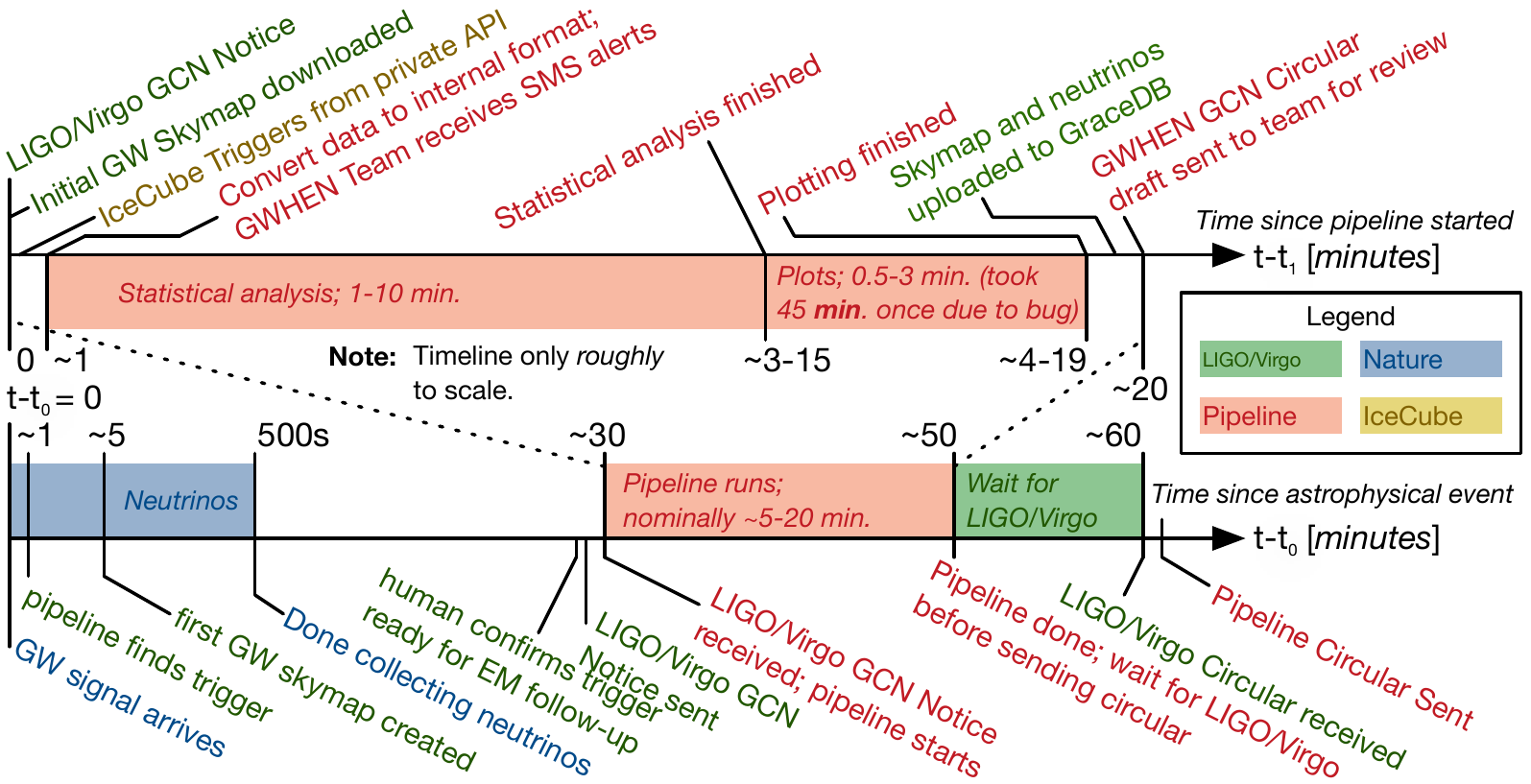}
    \caption{
        A typical timeline for a LIGO/Virgo trigger. The bottom timeline shows the
        full chain of events, particularly the role of LIGO/Virgo. The party
        responsible for each event or data product is color-coded according to
        the legend. The time of signal arrival at Earth is denoted $t_0$. Since
        \gwhen only ran on triggers marked by LIGO/Virgo as ready for EM followup
        (which required a human in the loop during O2 due to LIGO policy),
        there was a significant
        delay (usually 30 minutes) before processing began. The top timeline
        shows the detailed steps performed by \gwhen during a typical event and
        their associated delays. The time at which the pipeline starts
        processing a trigger is denoted $t_1$. The calculation of the joint
        likelihood $\mathcal{L}(\vec{x_s})$ and the plotting of the joint
        skymap typically took up the longest (and most variable) amount of
        time. Retrieving and disseminating data took negligible amounts of time
        in comparison. There were instances during O2 where edge cases,
        changing data formats, and bugs caused exceptionally long delays (often
        requiring human intervention), but these are considered atypical and
        not included in this timeline since their causes were easily fixed.
        Minute details of the pipeline's execution are also omitted from the
        timeline.
    }
    \label{fig:timeline}
\end{figure*}

During O1 and O2, there was typically a latency of about half an hour
(sometimes significantly more) between
LIGO/Virgo GW trigger identification and alert
dissemination via a GCN Notice
\cite{gcn} describing the GW event in VOEvent~\cite{voevent} format (see fig.
\ref{fig:timeline}). Most of this delay was due to the LIGO policy of human-in-the-loop
verification of trigger quality. After each VOEvent GCN Alert, a human-readable
GCN Circular would follow after a time period of half an hour or greater.

\Gwhen disseminated results in the form of human-readable GCN Circulars sent
out after the LIGO/Virgo GCN Circular. 
A GW+HEN GCN Circular was released after a LIGO/Virgo GCN Circular had been distributed in order to avoid confusing
EM partners and to add a third confirmation that the event was valid (in addition to the \emready GraceDB tag and GCN Notice).
This constraint allowed a comfortable target timeline
of half
an hour to run the combined analysis and human-in-the-loop checks
while still being ready to distribute our Circular as soon as the
LIGO/Virgo Circular was sent. By adhering to this timeline,
the search would not introduce any extra latency compared to the localization
provided by GW skymaps alone.

\subsection{\label{subsec:method}Data Analysis Method}

The analysis method used in this paper was an upgraded
version of Baret \etal~\cite{2012PhRvD..85j3004B}. A HEN event, provided by the IceCube Collabroation, was automatically internally marked as coincident with a GW if it
was detected within the $t_{\rm{GW}}\pm500$s time window (where $t_{\rm{GW}}$
is the time of the GW event) and if any region of the
GW's \ninetyCR had a neutrino signal likelihood density greater
than $10^{-4}$ deg$^{-2}$. A joint skymap plot was also generated for each
event and uploaded to the internal LIGO database (GraceDB). Interpretation
of observations was then performed by a human and described in an
automatically-generated GCN circular
for each event. Most of the terms required for a joint significance
calculation (described in \cite{2012PhRvD..85j3004B}) were calculated by \pipeline\ as
described below.
P-value calculations were not reported during O2.

\subsubsection{Coincidence Time Window}

The $t_{\rm{GW}}\pm500$s time window was chosen to account for upper limits on
observed GRB durations combined with central engine breakout time and
precursor delay (as described in \cite{2011APh....35....1B}). This time window makes no
further model assumptions and allows for \hen and \gw emissions in both the
precursor and GRB.

\subsubsection{Neutrino Data Processing}

In this study, we used a sample of through-going muons that were originally
designed for the IceCube gamma-ray follow-up (GFU)
program~\cite{icgfu2016,icecube-realtime} as described in Sec.~\ref{subsec:ic}.
We received these events in realtime with a latency of $\sim 1$min and accepted
any event reconstructed as an up-going neutrino as well as events that were
reconstructed as down-going neutrinos by applying a lower cut of 0.1 on their
BDT score ~\cite{ic-bdt-2014}.
The neutrino arrival time, direction, angular uncertainty, BDT
score, point spread function (PSF) and false alarm rate are used in our
correlation analysis. The PSF ($\nupsf$) is the probability distribution of
the neutrino source direction and can be explained with a Gaussian distribution
around the true source location ($\skyPosition$) as described in Eq. 6
of~\cite{2012PhRvD..85j3004B}. The HEN PSF term is incorporated in the likelihood
function of the joint GW+HEN analysis. We define the signal likelihood as
being equal to the PSF,
\begin{equation}
    \NeutrinoSignal = \nupsf,
\end{equation}
and the background likelihood is considered to be a flat distribution,
\begin{equation}
    \NeutrinoBackground = \frac{1}{2\pi}.
\end{equation}

\subsubsection{Gravitational Wave Data Processing}

The GW skymap $\GWSkymap$ gives the probability per $\rm{deg}^2$
that the GW came from sky direction $\skyPosition$. These
skymaps are calculated automatically by LIGO/Virgo in response to \GW event triggers
(\hyperref[subsec:data_sources]{discussed here}). For performance reasons,
probability was set to $0$ in regions outside the \ninetyCR of the
GW skymap (defined as the smallest region from which the
GW had a 90\% probability of originating), yielding a reduced
skymap $\reducedSkymap$. In the \pipeline\ code, pixels outside this region were
removed before the main analysis was run (effectively setting them to zero).

We defined the signal likelihood as being equal to the reduced skymap,
\begin{equation}
    \GWSignal = \reducedSkymap .
\end{equation}
Assuming that \GW triggers caused by background noise have an isotropic
distribution, the \GW background likelihood is simply
\begin{equation}
    \GWBackground = \frac{1}{4\pi}.
\end{equation}

\subsubsection{Joint Likelihood and Coincidence}

\pipeline\ calculated a joint likelihood ratio,
\begin{equation}
    \mathcal{L}(\vec{x_s}) = \frac{
        \GWSignal \NeutrinoSignal
    }{
        \GWBackground \NeutrinoBackground
    },
\end{equation}
for each temporally coincident neutrino. This is a modified version of the
formula described in \cite{2011APh....35....1B} without a joint p-value calculation.

There were two other features present in the codebase and method paper that
were not used during O2, namely, the neutrino clustering and galaxy catalog
features. The clustering code, which accounted for the possibility of multiple
neutrinos coming from the same source, was not used due to the low predicted multi-neutrino detection rate for O2. The
galaxy catalog code, which used information on galaxy locations to further
constrain source direction, was not used due to the lack of a galaxy catalog
whose range matched LIGO's maximum BNS detection range in O2.

\subsection{\label{subsec:implementation}Software Implementation}

\begin{figure}
    \begin{center}
        \resizebox{0.47\textwidth}{!}{\includegraphics{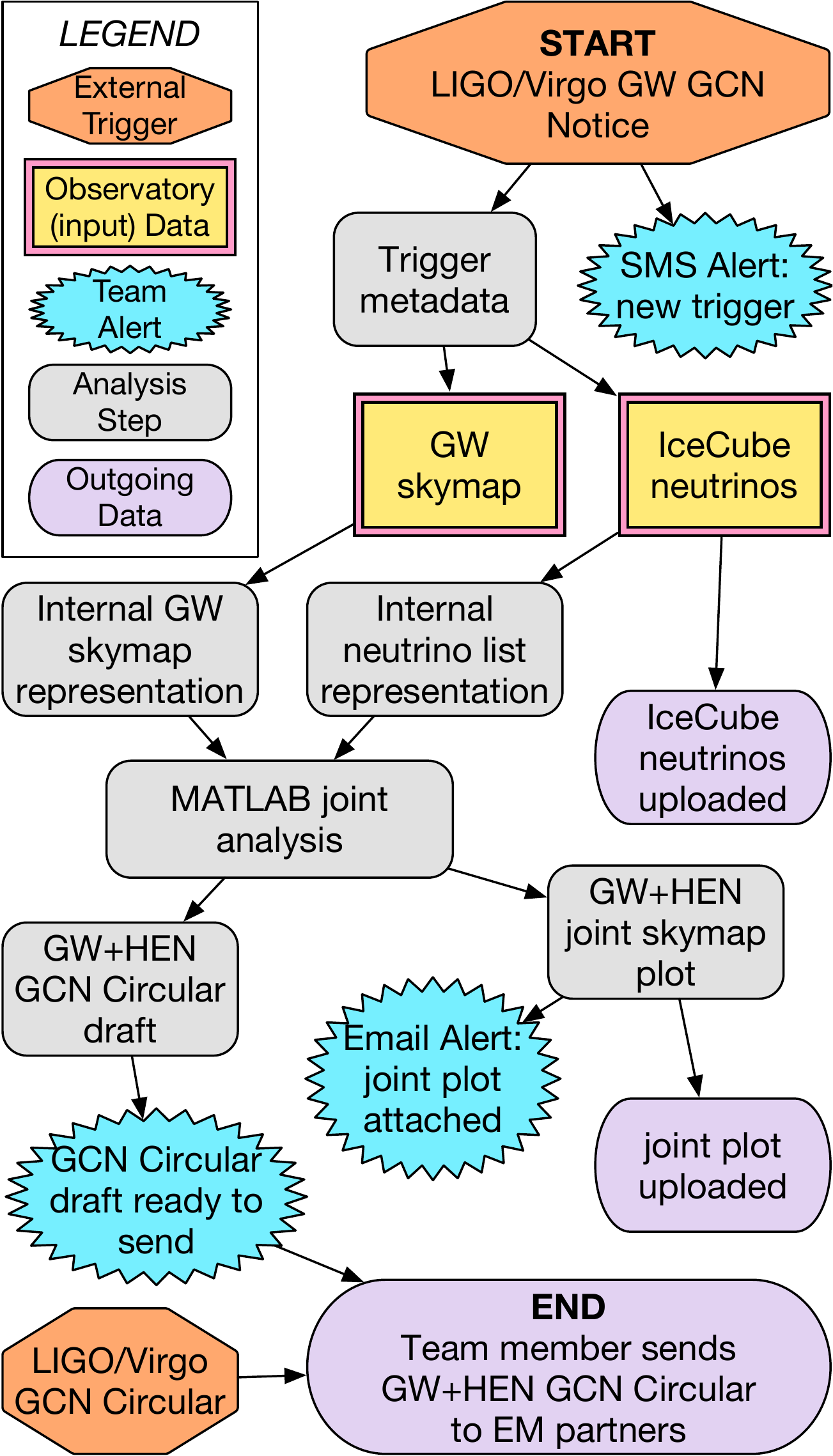}}
    \end{center}
    \caption{
        The pipeline's analysis steps as a DAG. Each node corresponds to an
        analysis output file (auxiliary steps have been omitted for clarity).
        \textit{External triggers} are created outside the pipeline but
        ingested and stored internally. The pipeline will repeatedly try
        to download \textit{observatory data} if not immediately available.
        \textit{Team alerts} are used internally. \textit{Outgoing data}
        are uploaded to GraceDB and \gwastro
        (except for the pipeline's GCN Circular, which is distributed to
        GCN after internal review).
    }
    \label{fig:dag}
\end{figure}

\subsubsection{Overview}
\label{subsubsec:overview}

\pipeline\ (Fig. \ref{fig:dag}) triggered on LIGO/Virgo GCN
Notices. Upon receiving a GCN Notice for a new LIGO/Virgo event,
the pipeline responded by creating an event directory on the analysis
server's filesystem for the new trigger. It
then pulled data from LIGO/Virgo (via GraceDB) and IceCube (via their
GFU API) as data became available; this included event metadata (in VOEvent
form) as well as actual skymaps and reconstructed parameters. The pipeline
would alert GW+HEN team members about the event; put input data into an
internal representation; run the analysis described in \ref{subsec:method};
upload results (neutrino triggers and a joint skymap plot) to GraceDB and
\gwastro; and, finally, send a joint skymap and GCN Circular draft to
team members, who would send out the GCN Circular once the LIGO/Virgo
GCN Circular had been distributed.

\subsubsection{Architecture}

The O2 pipeline was mostly implemented as a Python library (with the exception
of the analysis described in \ref{subsec:method},
which was a modified version of a reviewed MATLAB analysis codebase used in
previous offline searches, see e.g.,~\cite{2017ApJ...850L..35A,2017PhRvD..96b2005A,2016PhRvD..93l2010A,2014PhRvD..90j2002A,2012JPhCS.363a2022B,2012PhRvD..85j3004B,2009IJMPD..18.1655V,2008CQGra..25k4039A,2008APS..APR.E8006A,LIGOG060660,2018arXiv181010693A}.)
The steps in the pipeline (and their dependencies
on one another) are described in software by a Directed Acyclic Graph (DAG),
a simplified version of which is illustrated in Fig.~\ref{fig:dag}.
Two Python scripts running as daemon processes
(\texttt{gcnd} and \texttt{gwhend}, described below) ran
throughout O2 and automatically processed incoming events using
the tools provided by the Python library..

Trigger acquisition was the first step of the pipeline and was accomplished
by \texttt{gcnd}.
This daemon would parse incoming GCN notices in order to identify
new LIGO/Virgo \gw triggers. Once a suitable trigger was received,
\texttt{gcnd} would make a directory on the file system to hold data
associated with that trigger, save the VOEvent received from GCN in that
directory, and
extract GW trigger metadata into a new file (also in the trigger directory)
following \pipeline 's internal format. 
The pipeline is agnostic to trigger type and source and can accept triggers from
alternative sources using a suitable trigger listener alongside
(or instead of) \texttt{gcnd}. The combination with GWs was tested using a stream of realtime muon data provided by the IceCube Collaboration.

All subsequent steps of the pipeline were run automatically by
\texttt{gwhend},
which would monitor the directory where all triggers were stored
looking for partially-analyzed triggers and seeing if any further
steps of their analysis could be run. It accomplished this by
periodically checking whether any non-existing output files from the DAG
(Fig.~\ref{fig:dag}) had input files available and then generating
any such files. By iterating on this process, \texttt{gwhend} would
push the analysis for any recent event triggers as far as possible
while waiting for new data. Every other step described in
\ref{subsubsec:overview} and Fig.~\ref{fig:dag} (data analysis, team
alerts, and analysis output uploads) was run in this way
with the exception of the final GW+HEN GCN Circular submission.

This final step was performed manually by a team member both to guard
against spurious submission and to allow for tweaks to the GCN Circular
manuscript to satisfy LIGO requirements. The pipeline would generate a draft GW+HEN GCN Circular
and send it to team members, who could then modify it as needed
and email it to GCN for distribution once the LIGO/Virgo GCN
Circular (Sec.~\ref{subsec:timeline}) for that trigger had been
distributed.

\subsubsection{Features}

The pipeline was required to be highly extensible, reliable, and
reproducible.
The input dependencies and generation procedure for each node/file
in the DAG (Fig.~\ref{fig:dag}) were defined within a Python class
in the main \pipeline\ library.

Because each step in the procedure was self-contained with
explicitly-defined inputs, new analysis steps and procedures could
be added without affecting the overall stability of the pipeline.
Furthermore, having explicit dependencies and procedures made it
possible for \texttt{gwhend} to automatically execute steps of
the pipeline once data
became available (rather than executing them in a specific order),
recover gracefully from errors, and precisely define pipeline state
(in terms of existing vs. non-existing files) without introducing
a complex development framework.

The pipeline would also send out detailed stack traces and error
logs to the \pipeline\ team when exceptions occured. This, combined with
the independent nature of the analysis steps, proved useful at
several points during O2, e.g., when external sources provided 
wrongly-formatted skymap data or when external API changes caused 
exceptions while fetching data. In each case, the pipeline was 
able to continue processing unaffected parts
of each trigger's analysis while fixes were implemented,
minimizing delays.

Using a DAG also facilitated offline and manual analyses.
Analyses could be rerun with updated data without regenerating all
inputs or having to run the entire pipeline, and subcomponents of the
analysis could be manually run offline even if unneeded inputs were
missing.

The pipeline's ability to handle arbitrary trigger sources (once
trigger data is put into an internal format) will allow for both
new types of analysis as well as redundancy in trigger acquisition;
a more flexible \gw trigger-acquisition script using LIGO/Virgo's
LVAlert system was tested alongside \texttt{gcnd} during O2 to confirm
this.

In general, the flexibility and robustness of \pipeline\
will enable other planned upgrades
to the pipeline, including parallel file generation (for increased
performance), atomic step execution (for maximally robust fault
recovery and easier development), and pipeline state snapshots
(for rerunning analyses with different input data or analysis
procedures).

\section{\label{sec:conclusions}Conclusions}

\Gwhen established a robust platform for 
discoveries and enabled a low-latency GW/HEN search during
O2, ensuring that improved GW source localization could be provided in the
event of a joint neutrino detection and that joint-detection candidates could
be quickly found and analyzed.

No joint events were found during O2. Nonetheless, O2 provided a valuable
opportunity to implement, test, and refine \pipeline, laying the groundwork
for fast, reliable joint GW+HEN+GRB+MMA searches. 
during O2 have inspired important new features and performance improvements. 
Furthermore, the
proof-of-concept provided by \gwhen makes the framework a credibly reliable and
performant component of more complex multi-messenger search strategies in the
future.

The performance of the pipeline was within the specifications required to
ensure low-latency followup and was sufficiently fast to make the O2 pipeline
the fastest online multi-messenger search in the O2 era, with GW+HEN GCN
Circulars usually coming out immediately following the corresponding LIGO/Virgo GCN
Circulars. Nonetheless, there are numerous straightforward optimizations that
can further improve \pipeline\ performance by an order of magnitude. In
particular, less conservative pipeline triggering (by subscribing to trigger
updates directly from LIGO/Virgo) and optimized analysis code (using a faster
implementation of the analysis method) will enable this goal. 

\section{\label{sec:acknowledgments}Acknowledgments}

The authors thank the many researchers working in LIGO/Virgo, IceCube, and other
astrophysics projects that enabled this pipeline to operate successfully in O2.
In particular, they thank
Scott Barthelmy and Leo Singer of NASA for
providing helpful code and advice for working with GCN.

The authors are grateful to the IceCube Collaboration for providing neutrino dataset and support for using it for testing this algorithm. The Columbia Experimental Gravity group is grateful for the generous support of Columbia University in the City of New York and the National Science Foundation under grants PHY-1404462 and  PHY-1708028.
The authors are thankful for the generous support of the University of Florida and Columbia University in the City of New York.



\bibliographystyle{apsrev4-1}
%

\end{document}